\documentclass[twocolumn,showkeys,preprintnumbers,amsmath,amssymb]{revtex4}  

\usepackage{graphicx}
\usepackage{dcolumn}
\usepackage{bm}
\usepackage{multirow}
\usepackage{color}
\usepackage{hyperref}
\usepackage[normalem]{ulem}

\begin{document}

\title{Carrier and polarization dynamics in monolayer MoS$_2$}

\author{D. Lagarde$^1$}
\author{L. Bouet$^1$}
\author{X.~Marie$^{1}$}
\author{C.R.~Zhu$^2$}
\author{B.L.~Liu$^2$}
\author{T.~Amand$^1$}
\author{P.H.~Tan$^3$}
\author{B.~Urbaszek$^1$}
\affiliation{%
$^1$Universit\'e de Toulouse, INSA-CNRS-UPS, LPCNO, 135 Av. de Rangueil, 31077 Toulouse, France}
\affiliation{%
$^2$Beijing National Laboratory for Condensed Matter Physics, Institute of Physics, Chinese Academy of Sciences, Beijing 100190, China}
\affiliation{%
$^3$State Key Laboratory of Superlattices and Microstructures, Institute of Semiconductors, Chinese Academy of Sciences, Beijing,100083, China}


\begin{abstract}
In monolayer MoS$_2$ optical transitions across the direct bandgap are governed by chiral selection rules, allowing optical valley initialization. In time resolved photoluminescence (PL) experiments we find that both the polarization and emission dynamics do not change from 4K to 300K within our time resolution. We measure a high polarization and show that under pulsed excitation the emission polarization significantly decreases with increasing laser power. We find a fast exciton emission decay time on the order of 4ps. \textcolor{black}{The absence of a clear PL polarization decay within our time resolution suggests that the initially injected polarization dominates the steady state PL polarization. The observed decrease of the initial polarization with increasing pump photon energy hints at a possible ultrafast intervalley relaxation beyond the experimental ps time resolution.} By compensating the temperature induced change in bandgap energy with the excitation laser energy an emission polarization of 40\% is recovered at 300K, close to the maximum emission polarization for this sample at 4K. 

\end{abstract}

\maketitle

\section{Introduction}
\label{sec:intro}
\begin{figure}
\includegraphics[width=0.45\textwidth]{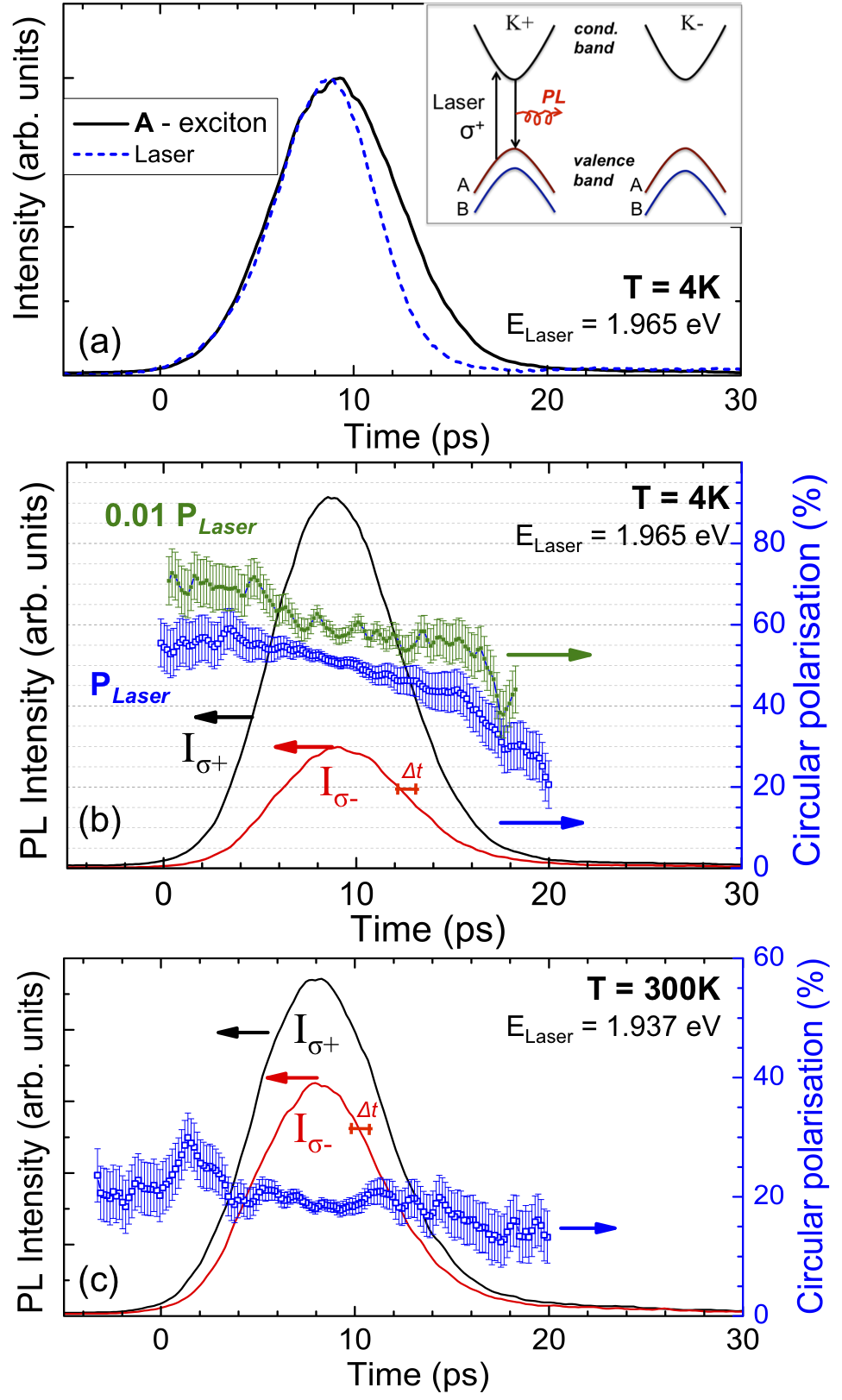}
\caption{\label{fig:fig1} \textbf{Time resolved photoluminescence of A-exciton.} (a) Laser pulse (blue line) and PL emission (black line) intensity at T = 4~K detected at maximum of A-exciton PL E$_{\text{Det}}=1.867$~eV as a function of time. Inset: Chiral optical selection rules in 1ML MoS$_2$ (b) T=4K, E$_{\text{Laser}}=1.965$~eV, E$_{\text{Det}}=1.867$~eV. Laser polarization $\sigma^+$. Left axis: $\sigma^+$ ($\sigma^-$) polarized PL emission intensity presented in black (red) as a function of time. Right axis:  Circular polarisation degree during exciton emission (blue hollow squares: excitation power P$_\text{Laser}\simeq550~\mu W/\mu m^2$, green full squares:0.01 P$_\text{Laser}$), errors bars take into account uncertainty in time origin $\Delta t\sim0.7ps$. (c) same as (b), but for T=300K, E$_{\text{Laser}}=1.937$~eV, E$_{\text{Det}}=1.828$~eV.}
\end{figure} 
Transition metal dichalcogenides such as MoS$_2$ emerge as an exciting class of atomically flat, two-dimensional materials for electronics \cite{Radisavljevic:2011a,Wang:2012a}, optics \cite{Kumar:2013a} and optoelectronics \cite{Sundaram:2013a}. In contrast to graphene, monolayer (ML) MoS$_2$ has a direct bandgap \cite{Mak:2010a,Splendiani:2010a} in the visible region of the optical spectrum. Inversion symmetry breaking (usually absent in graphene) together with the spin-orbit interaction leads to a unique coupling of carrier spin and k-space valley physics. The circular polarization ($\sigma^+$ or $\sigma^-$) of the absorbed or emitted photon can be directly associated with selective carrier excitation in one of the two non-equivalent K valleys ($K_+$ or $K_-$, respectively) \cite{Zhu:2011a,Cao:2012a,Xiao:2012a,Xiao:2013a}, where the role of strong excitonic effects merits further investigation in this context \cite{Cheiwchanchamnangij:2012a,Ramasubramaniam:2012a,Mak:2013a,Crowne:2013a}. The chiral optical selection rules open up very exciting possibilities of manipulating carriers in valleys with contrasting Berry phase curvatures \cite{Xiao:2010a}, aiming for experimental manifestations of the predicted valley Hall effect \cite{Xiao:2012a}. Also stable spin states have been predicted for valence and conduction states \cite{Ochoa:2013a,Wang:2013b} for this material. \\
\indent Up to now optical valley initialisation in ML MoS$_2$  is based on the analysis of the large circular polarization degree $P_c$ of the emitted light from the direct bandgap observed in continuous wave (cw) measurements following circularly polarized laser excitation \cite{Cao:2012a,Mak:2012a,Zeng:2012a,Sallen:2012a,Kioseoglou:2012a,Wu:2013a}. An important drawback seemed to be the drastic decrease of $P_c$ as the temperature is raised to 300K \cite{Mak:2012a,Zeng:2012a,Sallen:2012a,Wu:2013a}. In a simple approach, the stationary (time integrated) polarisation is determined by the initially created polarization $P_0$, the lifetime of the electron-hole pair $\tau$ and the polarization decay time $\tau_s$ through $P_c=P_0/(1+\tau/\tau_s)$ \cite{Meier:1984a}. We emphasize that the polarization decay time does not correspond directly to the carrier spin flip time as in most semiconductors like GaAs \cite{Meier:1984a}, but it includes the scattering time between the two non-equivalent K valleys (K$_+$ or K$_-$) \cite{Xiao:2012a}. \\
\indent In this letter we present the first time resolved polarization measurements in MoS$_2$ monolayers, providing vital information on the valley dynamics from 4K to room temperature. We determine the key parameters that govern the stationary polarization degree $P_c$: Using quasi-resonant excitation of the A-exciton transitions, we can infer that the photoluminescence (PL) decays within $\tau \simeq$4ps. For pulsed laser excitation, we observe a decrease of P$_c$ with increasing laser power. We show that the PL polarisation remains nearly constant in time for experiments from 4K up to 300K, a necessary condition for the success of future Valley Hall effect experiments based on optically initialized k-valley polarization \cite{Xiao:2012a}. In addition, $\tau$ does not vary significantly over this temperature range. These results are surprising when considering the reported decrease of $P_c$ in cw experiments when going from 4K to 300K reported in the literature \cite{Mak:2012a,Zeng:2012a,Sallen:2012a,Wu:2013a}. By tuning the laser following the shift of the A-exciton resonance with temperature we are able to recover at room temperature $\sim80\%$ of the polarization observed at 4K in our sample. \textcolor{black}{The absence of a clear PL polarization decay within our time resolution suggests that the initially injected polarization $P_0$, which dominates the steady state PL polarization, is responsible for this observation} 
\begin{figure}
\includegraphics[width=0.4\textwidth]{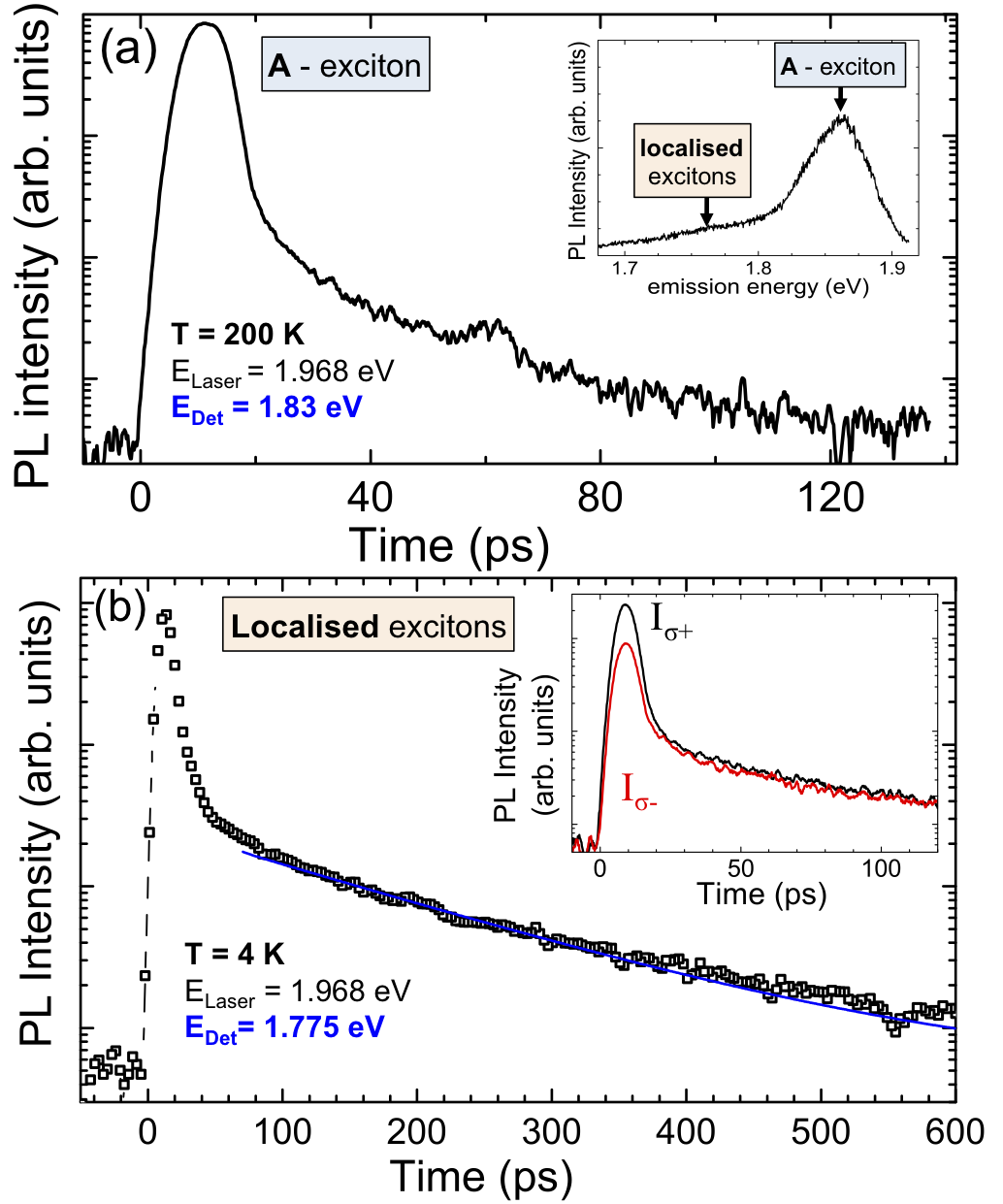}
\caption{\label{fig:fig2} \textbf{Time resolved photoluminescence} (a) T=200K, E$_{\text{Laser}}=1.968$~eV, E$_{\text{Det}}=1.83$~eV. \textbf{A-exciton } PL intensity as function of time, plotted in log-scale to emphasize weak, slow component that appears when raising T. The small peak at about 50ps is due to a laser reflection in the set-up. Inset: cw-PL spectrum at T=4K showing energy of intrinsic A-exciton emission and localised exciton emission. (b) T=4K, E$_{\text{Laser}}=1.968$~eV, E$_{\text{Det}}=1.775$~eV. \textbf{Localised exciton} PL intensity as a function of time (black). Blue line: fit with a single exponential decay ($\tau_{loc}\simeq125ps$). Inset: The polarization resolved emission of A-exciton and localised exciton which spectrally slightly overlap.
}
\end{figure} 

\section{Samples and Set-up}
\label{sec:methods}
MoS$_2$ flakes are obtained by micro-mechanical cleavage of a natural bulk MoS$_2$ crystal \cite{Novoselov:2005a}(from SPI Supplies, USA) on a Si/90 nm SiO$_2$ substrate. The 1ML region is identified by optical contrast and very clearly in PL spectroscopy \cite{Mak:2010a}. Experiments between T=4 and 300K are carried out in a confocal microscope optimized for polarized PL experiments\cite{Sallen:2011a}.  The MoS$_2$ flake is excited by picosecond pulses generated by a tunable frequency-doubled optical parametric oscillator (OPO) synchronously pumped by a mode-locked Ti:Sa laser. The typical pulse and spectral width are 1.6 ps and 3 meV  respectively; the repetition rate is 80 MHz. The laser wavelength can be tuned between 500 and 740 nm. The detection spot diameter is $\approx1$~$\mu$m. For time-integrated experiments, the PL emission is dispersed in a spectrometer and detected with a Si-CCD camera. For time-resolved experiments, the PL signal is dispersed by an imaging spectrometer and detected by a synchro-scan Hamamatsu Streak Camera  with an overall time resolution of 4 ps. The PL polarization $P_c$ defined as $P_c = \frac{I_{\sigma+}-I_{\sigma-}}{I_{\sigma+}+I_{\sigma-}}$ is analyzed by a quarter-wave plate placed in front of a linear polarizer. Here $I_{\sigma+}(I_{\sigma-})$ denotes the intensity of the right ($\sigma^+$) and left $(\sigma^-)$ circularly polarized emission. \\

\section{Experimental Results}
\label{sec:results}
\begin{figure}
\includegraphics[width=0.4\textwidth]{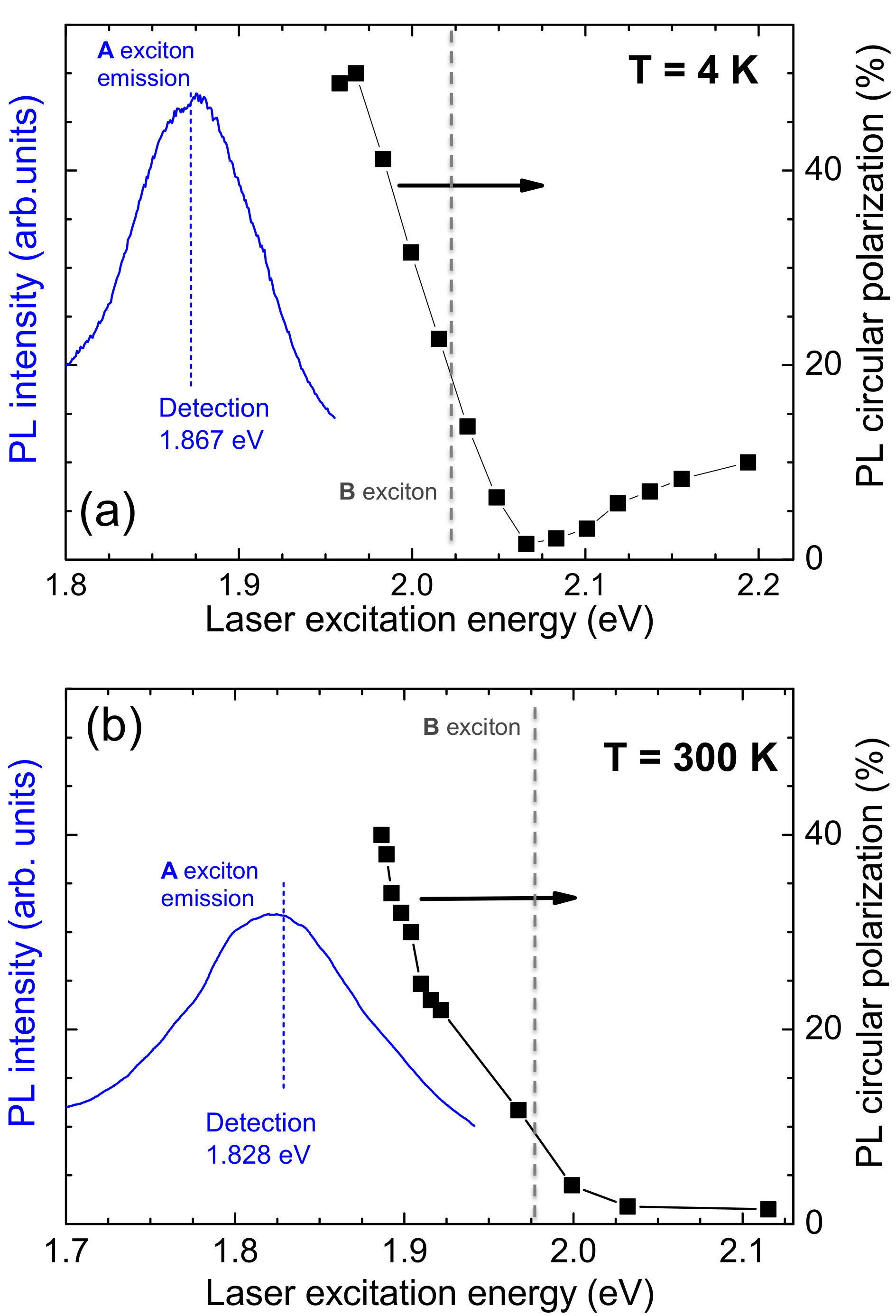}
\caption{\label{fig:fig3} \textbf{PL polarization as a function of laser excitation energy} (a) T = 4K, P$_\text{Laser}\simeq550~\mu W/\mu m^2$ detected on the A-exciton PL maximum E$_{\text{Det}}=1.867$~eV. The A-exciton emission (blue) is shown. (b) same as (a) but for T = 300 K, P$_\text{Laser}\simeq950~\mu W/\mu m^2$, E$_{\text{Det}}=1.828$~eV
}
\end{figure}

Figure \ref{fig:fig1}a displays the total photoluminescence intensity dynamics at T=4 K following a $\sigma^+$ polarized picosecond excitation laser pulse with an energy E$_{\text{Laser}}=1.965$~eV, which is within the broad A-exciton absorption line \cite{Mak:2012a,Mak:2013a}. The detection energy corresponds to the PL peak energy E$_{\text{Det}}=1.867$~eV.  The average laser power used for all experiments in figures \ref{fig:fig1} to \ref{fig:fig3} is below 1~$mW/\mu m^2$, well below absorption saturation and in the absence of sample heating effects, as discussed in detail in the supplementary material in section\ref{sec:Supp}. We do not observe any variation of the dynamics when the detection energy is varied within the A-exciton spectrum \cite{apa0}. Though the MoS$_2$ PL dynamics is very fast we see in figure \ref{fig:fig1} that it occurs on a slightly longer time scale compared to the one defined by the temporal resolution of the set-up (compare the MoS$_2$ PL and laser pulse detection in figure \ref{fig:fig1}a). Using a deconvolution based on Gaussian functions we can infer that the MoS$_2$ emission time is about 4.5 ps. We emphasize that this fast PL dynamics is obtained in excitation conditions where exclusively the A-exciton (not B) in the K$_+$ valley is excited (see inset of Fig.\ref{fig:fig1}a). Here the energy difference between excitation and detection $E_\text{Laser} - E_\text{Det}$ is about 100 meV.  Similarly fast dynamics were recorded in strongly non-resonant excitation conditions with $E_\text{Laser} - E_\text{Det} >1$~eV \cite{Korn:2011a,Shi:2013b}.\\
\indent In figure \ref{fig:fig1}b, the right ($I_{\sigma+}$) and left ($I_{\sigma-}$) circularly polarized luminescence components have been detected using a $\sigma^+$ polarized laser (measured $P_c^\text{Laser}>99\%$). Remarkably the PL circular polarization degree is large and remains almost constant during the short exciton emission, around 50\% for a laser excitation power P$_\text{Laser}\simeq550~\mu W/\mu m^2$ (blue curve in figure \ref{fig:fig1}b). Lowering the excitation power by two orders of magnitude has a strong impact: the polarization still remains nearly constant, but at a higher value of 60\% (green curve in figure \ref{fig:fig1}b). As $P_0<P_c^\text{Laser}$ either the polarization generation at this laser energy is not 100\% efficient (due to the optical selection rules) or \textcolor{black}{there exist an ultrafast, initial polarization decay due to intervalley relaxation much shorter than 1ps that we do not resolve.} Due to the sequential recording of $\sigma^+$ and $\sigma^-$ polarized kinetics there is an experimental uncertainty of $\Delta t\sim0.7ps$ when fixing the time origin of the $\sigma^+$ emission with respect to $\sigma^-$. This results in an experimental uncertainty when determining the circular PL polarization, as indicated by the error bars in figure \ref{fig:fig1}b. As a result of this time jitter and the short exciton emission time, our experiments do not allow an accurate determination of the spin/valley relaxation time $\tau_s$.
The emission at room temperature is very similar to 4K with $\tau\sim4.5ps$. Figure \ref{fig:fig1}c shows $P_c\simeq20\%$  constant in time for a laser excitation energy E$_{\text{Laser}}=1.937$~eV. The dependence of $P_c$ as a function of the excitation laser energy will be discussed below. Remarkably the PL circular polarization degree which probes the valley initialization measured in time-resolved experiment is similar to the one obtained in cw experiments as a result of the very short exciton lifetime and the absence of measurable polarization decay within this short emission time. \\
\indent In figure \ref{fig:fig2}a we show that when raising the temperature, in addition to the ps exciton decay a longer lived component is observed. This is similar to the findings of \textcite{Korn:2011a} under highly non-resonant excitation. This measured long lived PL component is essentially unpolarized in figure \ref{fig:fig2}a and the intensity is very weak compared to the short, main PL emission (log. scale). Therefore even at room temperature the short component, see figure \ref{fig:fig1}c, determines P$_c$. In addition to the main A-exciton also localised excitons emitting at lower energy (E$_{\text{Det}}=1.775$~eV) are observed \cite{Mak:2012a,Korn:2011a,Sallen:2012a}, see inset of figure \ref{fig:fig2}a. The localised exciton emission decays within about 125ps at T=4K (see figure \ref{fig:fig2}b) and is not detectable at higher temperature. The polarization dynamics is shown in the inset of figure \ref{fig:fig2}b: as the A-exciton and the broad, localised exciton emission spectrally slightly overlap, we detect the remaining A-exciton polarization at short times, before detecting the essentially unpolarized emission of the localised excitons. It is important to note that the localised exciton polarization dynamics is expected to be more sensitive to the sample parameters (substrate material, interface defects etc) than the A-exciton emission. \\

The time resolved measurements in figure \ref{fig:fig1}c show a nearly constant polarization during the exciton PL emission at T=300K. Yet in the literature the time-integrated polarization of the A-exciton PL at room temperature is reported to be considerably lower than at 4K \cite{Mak:2012a,Sallen:2012a,Wu:2013a}. We confirm these observations in our sample, see supplementary data in section\ref{sec:Supp}. Here it is important to take into account the considerable red-shift of the direct bandgap as the temperature increases \cite{Korn:2011a}. We have therefore performed time and polarization resolved photoluminescence excitation (PLE) experiments, see figure \ref{fig:fig3}, to vary the initially generated polarization P$_0$. The time-integrated PL is detected at its peak energy (1.867 eV and 1.828 eV at T=4 K and 300 K respectively). As already observed by different groups \cite{Mak:2012a,Zeng:2012a,Kioseoglou:2012a,Sallen:2012a} the PL circular polarization degree and thus the valley initialization decrease at T=4 K (figure \ref{fig:fig3}a) when the laser excitation energy increases : it varies for the sample investigated here from $P_c\simeq 50\%$ for $E_{\text{Laser}}=1.958$~eV down to a value close to zero for $E_{\text{Laser}}=2.06$~eV, similar to the findings by \textcite{Kioseoglou:2012a}. Here we observe that $P_c$ slightly increases again up to $P_c\simeq10\%$ for $E_{\text{Laser}}=2.2$~eV. Though the polarization minimum is observed roughly in the region where the B-exciton is photogenerated (see the vertical dotted line), its origin needs further clarification since the B-exciton absorbs and emits the same light helicity as the A-exciton in a given K valley \cite{Cao:2012a,Xiao:2012a}. The energetically close lying indirect transition from the $\Gamma$ valley valence band to the conduction band, which is unpolarized, could play a role \cite{Kormanyos:2013a,Wang:2013c}. When comparing results obtained on different samples in the literature \cite{Mak:2012a,Zeng:2012a,Sallen:2012a,Wu:2013a,Kioseoglou:2012a}, it is important to take into account the laser excitation (power, pulsed or cw) and the exact form of the emission spectrum, see section\ref{sec:Supp}.\\
\indent A key result is presented in figure \ref{fig:fig3}b. We perform the same PLE experiments as in figure \ref{fig:fig3}a, but at room temperature. When the laser is far from resonance, we observe close to zero polarization. Remarkably, as we lower the laser energy and become more and more resonant, the polarization drastically increases, in the same manner as at 4K. For the closest energy to resonance that was achievable in practice with our set-up (filter cut-off for stray laser light), we measure an emission polarization of $P_c(300K)=40\%$, close to the maximum observed at 4K of $P_c(4K)=50\%$ \cite{pol300K}. This is very encouraging as optical initialization of valley polarization with a suitable excitation source can therefore be very efficient even at room temperature. Here it would be extremely useful to investigate how  the strong Coulomb interaction \cite{Cheiwchanchamnangij:2012a,Ramasubramaniam:2012a,Mak:2013a,Crowne:2013a}  influences the polarization at 300K.

\section{Discussion}
\label{sec:Discussion} 
The few time resolved measurements reported in the literature for 1ML MoS$_2$ use highly non-resonant excitation and did not analyse the polarization of the emission \cite{Korn:2011a,Shi:2013b,Wang:2012b}. In \cite{Mak:2012a} and \cite{Shi:2013b}  the authors suggest that non-radiative recombination of the excitons could explain the first, short decay observed from 4K to 300K. We can infer from our time and polarization resolved measurements that the detected polarization in time-integrated experiments is due to the emission during this short time window. If this initial decay is indeed limited by non-radiative processes or an intrinsic exciton lifetime is still an open question. The exciton binding energy is estimated to be in the hundreds of meV range \cite{Cheiwchanchamnangij:2012a,Ramasubramaniam:2012a,Crowne:2013a}. Systems with large exciton binding energies such as organic films and carbon nanotubes have intrinsic exciton radiative lifetimes on the order of a few ps \cite{Varene:2012a,Perebeinos:2005a,Watanabe:1997a}. 
In our measurements  the PL decay time $\tau$ does not change with temperature within our time resolution, we do not observe any activation or any other typical signature of non-radiative processes. For comparison, for high quality GaAs quantum well structures the free Wannier-exciton radiative recombination time increases with temperature \cite{Feldmann:1987a}.

An argument in favour of a radiative exciton decay within a few ps comes from the observation of the B-exciton emission, 150meV higher in energy than the A-exciton \cite{Mak:2012a,Sallen:2012a}, as radiative recombination of the B-exciton is in competition with non-radiative decay and relaxation to the A-exciton. Here a theoretical prediction of the intrinsic exciton recombination is needed to guide future experiments in this promising system. \\
\indent In the future also the influence of the substrate on the polarization dynamics needs to be investigated. Here we used MoS$_2$ MLs on SiO$_2$, the most practical substrate for (opto-)electronic devices. Although time resolved experiments on suspended MoS$_2$ MLs gave similar time-resolved absorption results to experiments using substrates \cite{Shi:2013b}, a comparison with MoS$_2$ on BN where the localised exciton emission was suppressed \cite{Mak:2012a,Mak:2013a} would be important to clarify the nature of the observed PL emission.\\
\indent The very fast exciton decay time measured here in MoS$_2$ MLs could help explaining several key observations: (i) application of a transverse magnetic fields will only shown an influence on the polarization of the ML PL if the precession time is shorter than 4ps \cite{Sallen:2012a}. (ii) Coherence between valley excitons probed through the observation of stationary, linearly polarized luminescence in related WSe$_2$ devices, which are expected to have similar physical properties \cite{Jones:2013a}.

\section{Acknowledgements}
\label{sec:Acknowledgements}
We acknowledge partial funding from ERC Starting Grant No. 306719, Programme Investissements d'Avenir ANR-11-IDEX-0002-02, reference ANR-10-LABX-0037-NEXT, and CAS Grant No. 2011T1J37, and the National Basic Research Program of China (2009CB930502, 2009CB929301) and National Science Foundation of China [Grants No. 11174338, No. 11225421, No. 10934007 and No. 10911130356 (Spinman)]. Delphine Lagarde and Louis Bouet contributed equally to this work.

\section{Supplementary Material}
\label{sec:Supp}
\setcounter{figure}{0}

In this supplementary material we present additional, power dependent results obtained under pulsed and a continuous wave excitation conditions at 4K and room temperature in a MoS$_2$ monolayer sample. 

\begin{figure}
\includegraphics[width=0.4\textwidth]{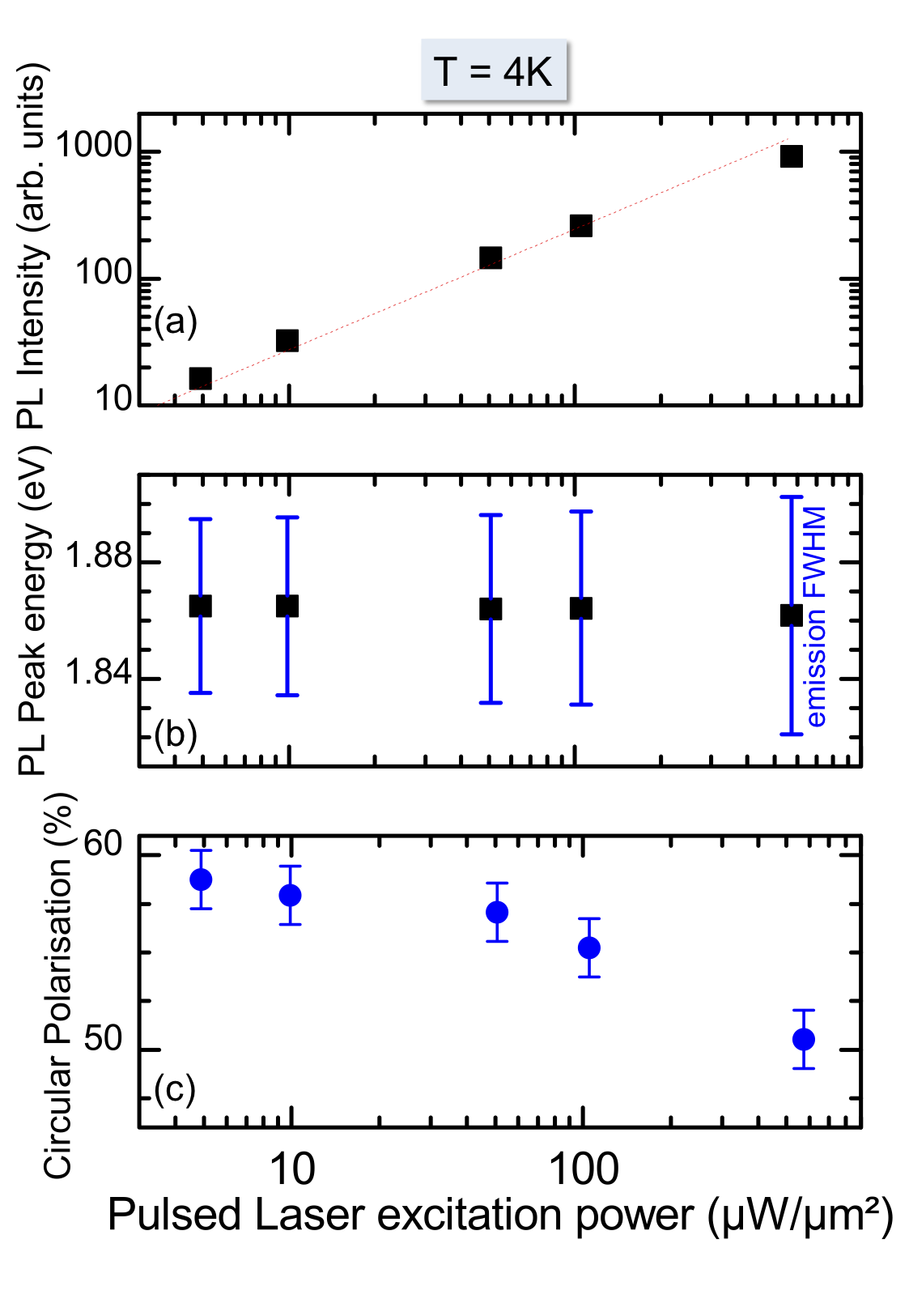}
\renewcommand{\thefigure}{S\arabic{figure}}
\caption{\label{fig:figsi1} \textbf{Power dependence of Photoluminescence under pulsed Laser excitation at 4K} using streak camera detection at E$_{\text{Det}}=1.867$~eV and E$_{\text{Laser}}=1.965$~eV (a) PL intensity as a function of laser power (b) Peak position of A-exciton PL emission (black squares). The FWHM of the transitions is indicated (blue lines). (c) PL circular polarization degree detected at central point of PL kinetics as a function of laser power (blue circles).}
\end{figure} 
Optical spectroscopy of MoS$_2$ monolayers is carried out either with standard continuous wave (cw) lasers such as the HeNe laser (1.96eV) or with tunable pulsed lasers such as a ps frequency-doubled optical parametric oscillator synchronously pumped by a mode-locked Ti:Sa laser (OPO), see main text and the work by \textcite{Kioseoglou:2012a}. The laser power used in the experiments is dictated by practical considerations: at low power the signal level has to be sufficient to achieve a high signal-to-noise ratio, at high excitation power structural damage of the MoS$_2$ monolayer has to be avoided. To the best of our knowledge their is no systematic study of the power dependence of the circular polarization degree $P_c$ of the photoluminescence of ML MoS$_2$. In the experiments below we demonstrate that the laser excitation power range used for the results presented in the main text is still within the linear absorption regime (below saturation) and does not induce any measurable sample heating or degradation.\\
\subsection{Experiments using pulsed Laser excitation}
The power of the OPO used in the experiments in the main text was chosen to provide high signal-to-noise ratio time resolved photoluminescence (TRPL) spectra without causing sample damage. In general, time resolved measurements using a streak-camera demand a higher signal level than time-integrated measurements using a sensitive Si-CCD camera. In figure \ref{fig:figsi1}(a) we show the PL intensity as a function of laser power over 2 orders of magnitude detected with the streak camera system i.e. under the same conditions as the data in figure 1 of the main text. The absorption is not saturated for the laser power values used in time resolved spectroscopy, see figure 1 of the main text, where P$_\text{Laser}\simeq550~\mu W/\mu m^2$ corresponds to the highest power used in figure S1. 

\begin{figure}
\includegraphics[width=0.4\textwidth]{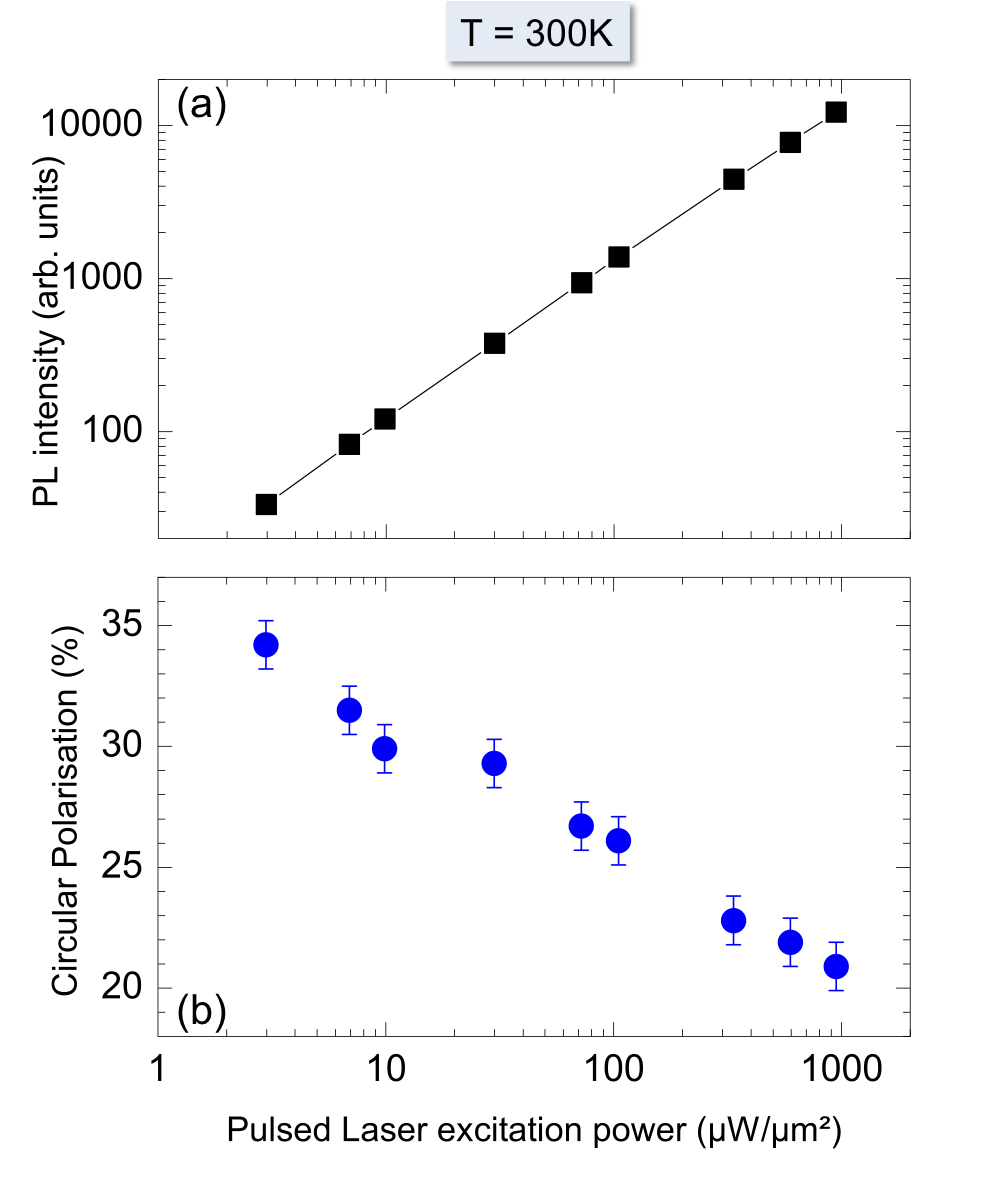}
\renewcommand{\thefigure}{S\arabic{figure}}
\caption{\label{fig:figsi2} \textbf{Power dependence of Photoluminescence under pulsed Laser excitation at 300K} detected at Si-CCDfor and E$_{\text{Laser}}=1.913$~eV (a) PL intensity as a function of laser power (black squares) (b) PL circular polarization degree detected at E$_{\text{Det}}=1.828$~eV as a function of laser power (blue circles).}
\end{figure}
\begin{figure}
\includegraphics[width=0.4\textwidth]{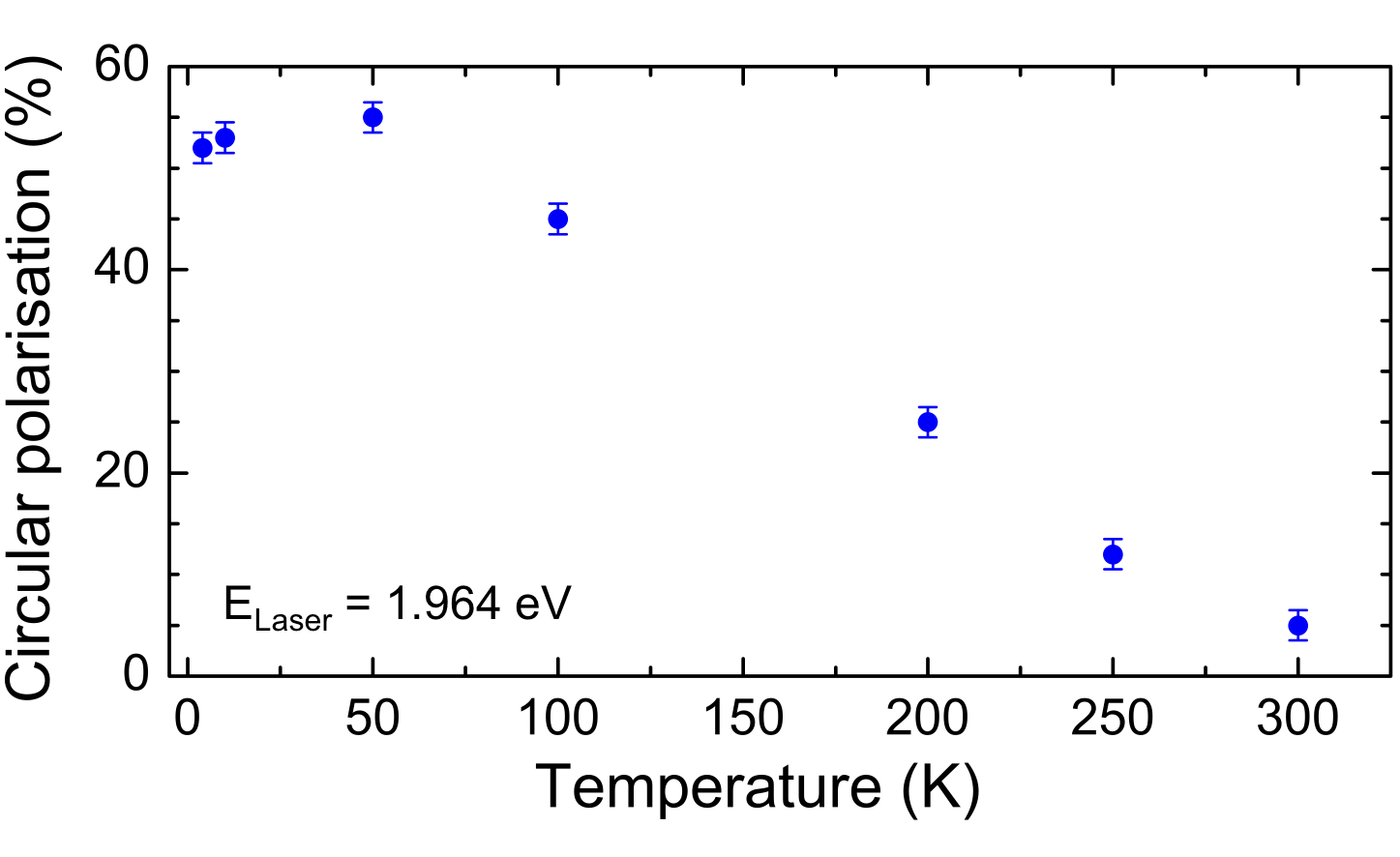}
\renewcommand{\thefigure}{S\arabic{figure}}
\caption{\label{fig:figsi3} A-exction PL polarisation degree detected at PL maximum as a function of temperature for fixed excitation laser energy E$_{\text{Laser}}=1.964$~eV.
}
\end{figure}

We find as a general trend that the PL polarization decreases as the laser power increases, a shown in figure \ref{fig:figsi1}(c). The transition energy is constant for the laser power range investigated, see figure \ref{fig:figsi1}(b). We therefore conclude that sample lattice heating is an unlikely cause of this polarization decrease, it is possible that depolarization due to a Dyakonov-Perel type mechanism plays a role.\\
At a temperature of 300K the optical emission of the MoS$_2$ monolayer is weaker than at 4K. In order to obtain an acceptable signal to noise ratio for the PL at low laser power, we performed the power dependent studies at 300K using a sensitive Si-CCD camera at a detection energy of 1.828~eV for a laser energy of 1.913~eV. In figure \ref{fig:figsi2}(a) we confirm that also at 300K we are still working in the linear absorption regime. In figure \ref{fig:figsi1}(b) we show a decrease of the circular PL polarization degree as the laser power increases, similar to the trend shown at low temperature. \\
As previously reported in the literature \cite{Mak:2012a,Zeng:2012a,Sallen:2012a,Wu:2013a}, we observe a decrease in our sample of the circular PL polarization as a function of temperature, see figure \ref{fig:figsi3}. Here the laser energy is kept constant and does not follow the bandgap shrinking with temperature.\\
\begin{figure}
\includegraphics[width=0.45\textwidth]{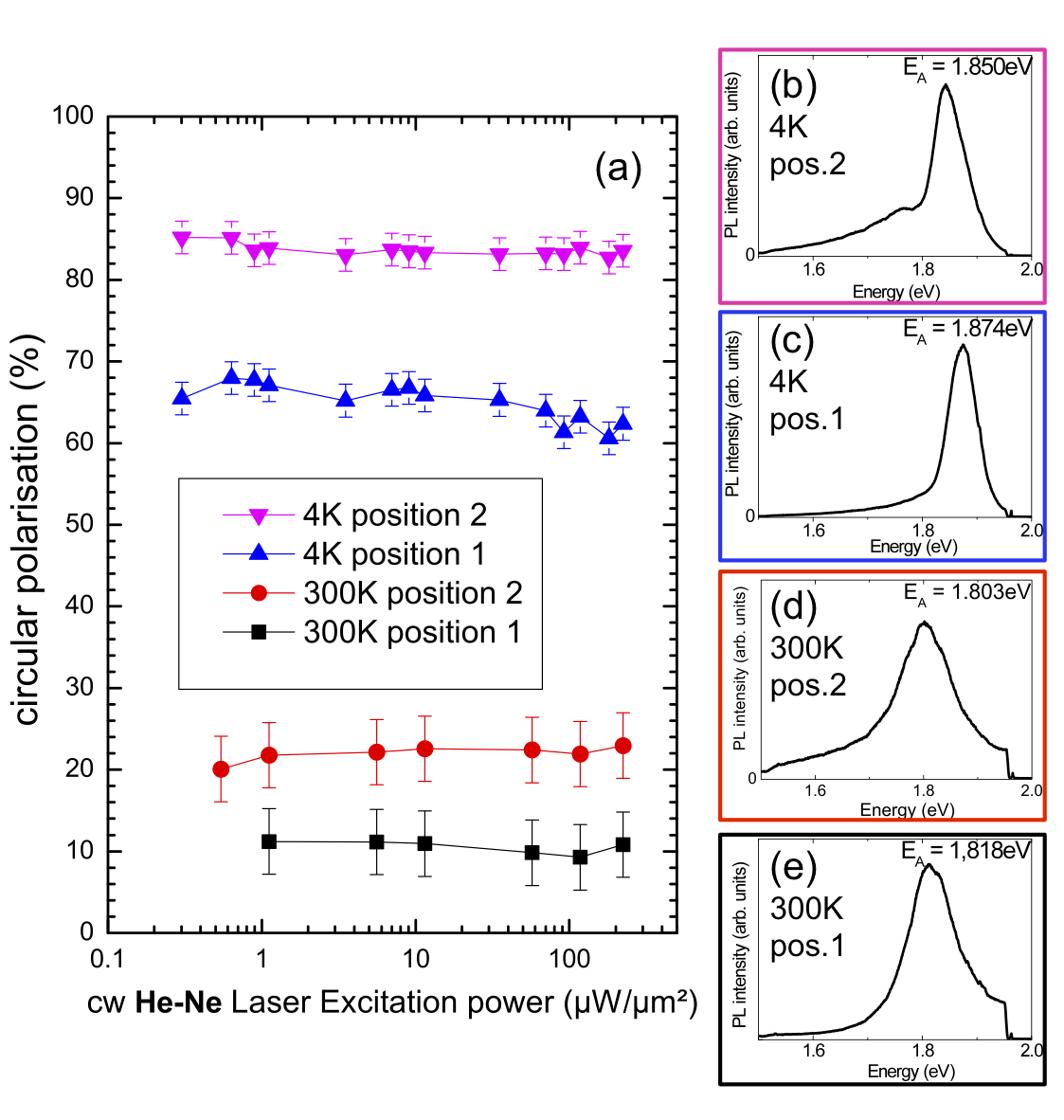}
\renewcommand{\thefigure}{S\arabic{figure}}
\caption{\label{fig:figsi4} \textbf{Photoluminescence of monolayer MoS$_2$ using cw HeNe Laser} (a) Polarization of the PL emission at 300K: black squares from \textit{position 1} on the sample, red circles from \textit{position 2}; at 4K:blue upwards triangles \textit{position 1}, pink downwards triangles \textit{position 2}. (b) spectrum of PL at 4K at \textit{position 2}, the energy of the maximum A-exciton emission is E$_A$ (c)  PL at 4K at \textit{position 1} (d)  PL at 300K at \textit{position 2} (e)  PL at 300K at \textit{position 1}}
\end{figure} 

\subsection{Experiments using cw Helium-Neon laser excitation}
Very high PL polarization under cw HeNe laser excitation at 4K and at 300K has been reported before \cite{Mak:2012a,Sallen:2012a} and is confirmed in figure \ref{fig:figsi4}(a). First, the laser emission is spectrally narrower than the pulsed OPO. Second, although the average laser power monitored with a photodiode, which is used to measure the laser power plotted in figure S1 and S2, might be comparable in the experiments, the  power is several orders of magnitude lower than the peak power in pulsed operation. Here we use the same set-up as in our previous work \cite{Sallen:2012a}, but a different sample. Over the investigated power range the polarization does not change drastically, in contrast to the measurements with the OPO excitation. This seems to indicate that the generation rate with the OPO at the lowest power is still much higher than the HeNe generation rate at its maximum. \\
Our confocal microscope is also used for spectroscopy on single semiconductor quantum dots (typical diameter 20nm) \cite{Urbaszek:2013a} and due the high spatial resolution we are able to investigate \textit{different} regions of the \textit{same} MoS$_2$ monolayer as the sample is mounted on piezo-positioners with a step-size in nm-range in a variable temperature cryostat. We find measurable variations of the polarization for different positions in the monolayer: The first type of spectra have a more pronounced A-exciton emission and only a very small contribution of the localized exciton. Analysing the A-exciton emission, we find $P_c$ on the order of 70\% at 4K and 10\% at 300K, see figure \ref{fig:figsi3}. The second type of spectra is about 15 meV red-shifted with a more pronounced contribution of  localized excitons. Here the A-exciton emission is more polarized, with $P_c$ on the order of 85\% at 4K and 25\% at 300K, close to the values reported in \cite{Sallen:2012a}. Please note that we cannot directly compare the PL polarization measured in figures \ref{fig:figsi1} and \ref{fig:figsi2} (OPO excitation) with the results in figure \ref{fig:figsi4} (HeNe excitation at 1.96eV) as the laser energies are different and the polarization changes abruptly for different energies (see main text).


\begin{thebibliography}{38}
\expandafter\ifx\csname natexlab\endcsname\relax\def\natexlab#1{#1}\fi
\expandafter\ifx\csname bibnamefont\endcsname\relax
  \def\bibnamefont#1{#1}\fi
\expandafter\ifx\csname bibfnamefont\endcsname\relax
  \def\bibfnamefont#1{#1}\fi
\expandafter\ifx\csname citenamefont\endcsname\relax
  \def\citenamefont#1{#1}\fi
\expandafter\ifx\csname url\endcsname\relax
  \def\url#1{\texttt{#1}}\fi
\expandafter\ifx\csname urlprefix\endcsname\relax\def\urlprefix{URL }\fi
\providecommand{\bibinfo}[2]{#2}
\providecommand{\eprint}[2][]{\url{#2}}

\bibitem[{\citenamefont{Radisavljevic et~al.}(2011)\citenamefont{Radisavljevic,
  Radenovic, Brivio, Giacometti, and Kis}}]{Radisavljevic:2011a}
\bibinfo{author}{\bibfnamefont{B.}~\bibnamefont{Radisavljevic}},
  \bibinfo{author}{\bibfnamefont{A.}~\bibnamefont{Radenovic}},
  \bibinfo{author}{\bibfnamefont{J.}~\bibnamefont{Brivio}},
  \bibinfo{author}{\bibfnamefont{V.}~\bibnamefont{Giacometti}},
  \bibnamefont{and} \bibinfo{author}{\bibfnamefont{A.}~\bibnamefont{Kis}},
  \bibinfo{journal}{Nature. Nanotech.} \textbf{\bibinfo{volume}{6}},
  \bibinfo{pages}{147} (\bibinfo{year}{2011}).

\bibitem[{\citenamefont{Wang et~al.}(2012{\natexlab{a}})\citenamefont{Wang, Yu,
  Lee, Fang, Hsu, Herring, Chin, Dubey, Li, Kong et~al.}}]{Wang:2012a}
\bibinfo{author}{\bibfnamefont{H.}~\bibnamefont{Wang}},
  \bibinfo{author}{\bibfnamefont{L.}~\bibnamefont{Yu}},
  \bibinfo{author}{\bibfnamefont{Y.}~\bibnamefont{Lee}},
  \bibinfo{author}{\bibfnamefont{W.}~\bibnamefont{Fang}},
  \bibinfo{author}{\bibfnamefont{A.}~\bibnamefont{Hsu}},
  \bibinfo{author}{\bibfnamefont{P.}~\bibnamefont{Herring}},
  \bibinfo{author}{\bibfnamefont{M.}~\bibnamefont{Chin}},
  \bibinfo{author}{\bibfnamefont{M.}~\bibnamefont{Dubey}},
  \bibinfo{author}{\bibfnamefont{L.}~\bibnamefont{Li}},
  \bibinfo{author}{\bibfnamefont{J.}~\bibnamefont{Kong}}, \bibnamefont{et~al.},
  in \emph{\bibinfo{booktitle}{Electron Devices Meeting (IEDM), 2012 IEEE
  International}} (\bibinfo{year}{2012}{\natexlab{a}}), pp.
  \bibinfo{pages}{4.6.1--4.6.4}, ISSN \bibinfo{issn}{0163-1918}.

\bibitem[{\citenamefont{Kumar et~al.}(2013)\citenamefont{Kumar, Najmaei, Cui,
  Ceballos, Ajayan, Lou, and Zhao}}]{Kumar:2013a}
\bibinfo{author}{\bibfnamefont{N.}~\bibnamefont{Kumar}},
  \bibinfo{author}{\bibfnamefont{S.}~\bibnamefont{Najmaei}},
  \bibinfo{author}{\bibfnamefont{Q.}~\bibnamefont{Cui}},
  \bibinfo{author}{\bibfnamefont{F.}~\bibnamefont{Ceballos}},
  \bibinfo{author}{\bibfnamefont{P.~M.} \bibnamefont{Ajayan}},
  \bibinfo{author}{\bibfnamefont{J.}~\bibnamefont{Lou}}, \bibnamefont{and}
  \bibinfo{author}{\bibfnamefont{H.}~\bibnamefont{Zhao}},
  \bibinfo{journal}{Phys. Rev. B} \textbf{\bibinfo{volume}{87}},
  \bibinfo{pages}{161403} (\bibinfo{year}{2013}),
  \urlprefix\url{http://link.aps.org/doi/10.1103/PhysRevB.87.161403}.

\bibitem[{\citenamefont{Sundaram et~al.}(2013)\citenamefont{Sundaram, Engel,
  Lombardo, Krupke, Ferrari, Avouris, and Steiner}}]{Sundaram:2013a}
\bibinfo{author}{\bibfnamefont{R.~S.} \bibnamefont{Sundaram}},
  \bibinfo{author}{\bibfnamefont{M.}~\bibnamefont{Engel}},
  \bibinfo{author}{\bibfnamefont{A.}~\bibnamefont{Lombardo}},
  \bibinfo{author}{\bibfnamefont{R.}~\bibnamefont{Krupke}},
  \bibinfo{author}{\bibfnamefont{A.~C.} \bibnamefont{Ferrari}},
  \bibinfo{author}{\bibfnamefont{P.}~\bibnamefont{Avouris}}, \bibnamefont{and}
  \bibinfo{author}{\bibfnamefont{M.}~\bibnamefont{Steiner}},
  \bibinfo{journal}{Nano Letters} \textbf{\bibinfo{volume}{13}},
  \bibinfo{pages}{1416} (\bibinfo{year}{2013}),
  \eprint{http://pubs.acs.org/doi/pdf/10.1021/nl400516a},
  \urlprefix\url{http://pubs.acs.org/doi/abs/10.1021/nl400516a}.

\bibitem[{\citenamefont{Mak et~al.}(2010)\citenamefont{Mak, Lee, Hone, Shan,
  and Heinz}}]{Mak:2010a}
\bibinfo{author}{\bibfnamefont{K.~F.} \bibnamefont{Mak}},
  \bibinfo{author}{\bibfnamefont{C.}~\bibnamefont{Lee}},
  \bibinfo{author}{\bibfnamefont{J.}~\bibnamefont{Hone}},
  \bibinfo{author}{\bibfnamefont{J.}~\bibnamefont{Shan}}, \bibnamefont{and}
  \bibinfo{author}{\bibfnamefont{T.~F.} \bibnamefont{Heinz}},
  \bibinfo{journal}{Phys. Rev. Lett.} \textbf{\bibinfo{volume}{105}},
  \bibinfo{pages}{136805} (\bibinfo{year}{2010}).

\bibitem[{\citenamefont{Splendiani et~al.}(2010)\citenamefont{Splendiani, Sun,
  Zhang, Li, Kim, Chim, Galli, and Wang}}]{Splendiani:2010a}
\bibinfo{author}{\bibfnamefont{A.}~\bibnamefont{Splendiani}},
  \bibinfo{author}{\bibfnamefont{L.}~\bibnamefont{Sun}},
  \bibinfo{author}{\bibfnamefont{Y.}~\bibnamefont{Zhang}},
  \bibinfo{author}{\bibfnamefont{T.}~\bibnamefont{Li}},
  \bibinfo{author}{\bibfnamefont{J.}~\bibnamefont{Kim}},
  \bibinfo{author}{\bibfnamefont{C.-Y.} \bibnamefont{Chim}},
  \bibinfo{author}{\bibfnamefont{G.}~\bibnamefont{Galli}}, \bibnamefont{and}
  \bibinfo{author}{\bibfnamefont{F.}~\bibnamefont{Wang}},
  \bibinfo{journal}{Nano Letters} \textbf{\bibinfo{volume}{10}},
  \bibinfo{pages}{1271} (\bibinfo{year}{2010}).

\bibitem[{\citenamefont{Zhu et~al.}(2011)\citenamefont{Zhu, Cheng, and
  Schwingenschl\"ogl}}]{Zhu:2011a}
\bibinfo{author}{\bibfnamefont{Z.~Y.} \bibnamefont{Zhu}},
  \bibinfo{author}{\bibfnamefont{Y.~C.} \bibnamefont{Cheng}}, \bibnamefont{and}
  \bibinfo{author}{\bibfnamefont{U.}~\bibnamefont{Schwingenschl\"ogl}},
  \bibinfo{journal}{Phys. Rev. B} \textbf{\bibinfo{volume}{84}},
  \bibinfo{pages}{153402} (\bibinfo{year}{2011}).

\bibitem[{\citenamefont{Cao et~al.}(2012)\citenamefont{Cao, Wang, Han, Ye, Zhu,
  Shi, Niu, Tan, Wang, Liu et~al.}}]{Cao:2012a}
\bibinfo{author}{\bibfnamefont{T.}~\bibnamefont{Cao}},
  \bibinfo{author}{\bibfnamefont{G.}~\bibnamefont{Wang}},
  \bibinfo{author}{\bibfnamefont{W.}~\bibnamefont{Han}},
  \bibinfo{author}{\bibfnamefont{H.}~\bibnamefont{Ye}},
  \bibinfo{author}{\bibfnamefont{C.}~\bibnamefont{Zhu}},
  \bibinfo{author}{\bibfnamefont{J.}~\bibnamefont{Shi}},
  \bibinfo{author}{\bibfnamefont{Q.}~\bibnamefont{Niu}},
  \bibinfo{author}{\bibfnamefont{P.}~\bibnamefont{Tan}},
  \bibinfo{author}{\bibfnamefont{E.}~\bibnamefont{Wang}},
  \bibinfo{author}{\bibfnamefont{B.}~\bibnamefont{Liu}}, \bibnamefont{et~al.},
  \bibinfo{journal}{Nature Communications} \textbf{\bibinfo{volume}{3}},
  \bibinfo{pages}{887} (\bibinfo{year}{2012}).

\bibitem[{\citenamefont{Xiao et~al.}(2012)\citenamefont{Xiao, Liu, Feng, Xu,
  and Yao}}]{Xiao:2012a}
\bibinfo{author}{\bibfnamefont{D.}~\bibnamefont{Xiao}},
  \bibinfo{author}{\bibfnamefont{G.-B.} \bibnamefont{Liu}},
  \bibinfo{author}{\bibfnamefont{W.}~\bibnamefont{Feng}},
  \bibinfo{author}{\bibfnamefont{X.}~\bibnamefont{Xu}}, \bibnamefont{and}
  \bibinfo{author}{\bibfnamefont{W.}~\bibnamefont{Yao}},
  \bibinfo{journal}{Phys. Rev. Lett.} \textbf{\bibinfo{volume}{108}},
  \bibinfo{pages}{196802} (\bibinfo{year}{2012}).

\bibitem[{\citenamefont{Li et~al.}(2013)\citenamefont{Li, Zhang, and
  Niu}}]{Xiao:2013a}
\bibinfo{author}{\bibfnamefont{X.}~\bibnamefont{Li}},
  \bibinfo{author}{\bibfnamefont{F.}~\bibnamefont{Zhang}}, \bibnamefont{and}
  \bibinfo{author}{\bibfnamefont{Q.}~\bibnamefont{Niu}},
  \bibinfo{journal}{Phys. Rev. Lett.} \textbf{\bibinfo{volume}{110}},
  \bibinfo{pages}{066803} (\bibinfo{year}{2013}),
  \urlprefix\url{http://link.aps.org/doi/10.1103/PhysRevLett.110.066803}.

\bibitem[{\citenamefont{Cheiwchanchamnangij and
  Lambrecht}(2012)}]{Cheiwchanchamnangij:2012a}
\bibinfo{author}{\bibfnamefont{T.}~\bibnamefont{Cheiwchanchamnangij}}
  \bibnamefont{and} \bibinfo{author}{\bibfnamefont{W.~R.~L.}
  \bibnamefont{Lambrecht}}, \bibinfo{journal}{Phys. Rev. B}
  \textbf{\bibinfo{volume}{85}}, \bibinfo{pages}{205302}
  (\bibinfo{year}{2012}),
  \urlprefix\url{http://link.aps.org/doi/10.1103/PhysRevB.85.205302}.

\bibitem[{\citenamefont{Ramasubramaniam}(2012)}]{Ramasubramaniam:2012a}
\bibinfo{author}{\bibfnamefont{A.}~\bibnamefont{Ramasubramaniam}},
  \bibinfo{journal}{Phys. Rev. B} \textbf{\bibinfo{volume}{86}},
  \bibinfo{pages}{115409} (\bibinfo{year}{2012}),
  \urlprefix\url{http://link.aps.org/doi/10.1103/PhysRevB.86.115409}.

\bibitem[{\citenamefont{Mak et~al.}(2013)\citenamefont{Mak, He, Changgu, Lee,
  Hone, Heinz, and Shan}}]{Mak:2013a}
\bibinfo{author}{\bibfnamefont{K.~F.} \bibnamefont{Mak}},
  \bibinfo{author}{\bibfnamefont{K.}~\bibnamefont{He}},
  \bibinfo{author}{\bibnamefont{Changgu}},
  \bibinfo{author}{\bibfnamefont{G.~H.} \bibnamefont{Lee}},
  \bibinfo{author}{\bibfnamefont{J.}~\bibnamefont{Hone}},
  \bibinfo{author}{\bibfnamefont{T.~F.} \bibnamefont{Heinz}}, \bibnamefont{and}
  \bibinfo{author}{\bibfnamefont{J.}~\bibnamefont{Shan}},
  \bibinfo{journal}{Nature Materials} \textbf{\bibinfo{volume}{12}},
  \bibinfo{pages}{207} (\bibinfo{year}{2013}).

\bibitem[{\citenamefont{Crowne et~al.}(2013)\citenamefont{Crowne, Amani,
  Birdwell, Chin, O'Regan, Najmaei, Liu, Ajayan, Lou, and
  Dubey}}]{Crowne:2013a}
\bibinfo{author}{\bibfnamefont{F.~J.} \bibnamefont{Crowne}},
  \bibinfo{author}{\bibfnamefont{M.}~\bibnamefont{Amani}},
  \bibinfo{author}{\bibfnamefont{A.~G.} \bibnamefont{Birdwell}},
  \bibinfo{author}{\bibfnamefont{M.~L.} \bibnamefont{Chin}},
  \bibinfo{author}{\bibfnamefont{T.~P.} \bibnamefont{O'Regan}},
  \bibinfo{author}{\bibfnamefont{S.}~\bibnamefont{Najmaei}},
  \bibinfo{author}{\bibfnamefont{Z.}~\bibnamefont{Liu}},
  \bibinfo{author}{\bibfnamefont{P.~M.} \bibnamefont{Ajayan}},
  \bibinfo{author}{\bibfnamefont{J.}~\bibnamefont{Lou}}, \bibnamefont{and}
  \bibinfo{author}{\bibfnamefont{M.}~\bibnamefont{Dubey}},
  \bibinfo{journal}{Phys. Rev. B} \textbf{\bibinfo{volume}{88}},
  \bibinfo{pages}{235302} (\bibinfo{year}{2013}),
  \urlprefix\url{http://link.aps.org/doi/10.1103/PhysRevB.88.235302}.

\bibitem[{\citenamefont{Xiao et~al.}(2010)\citenamefont{Xiao, Chang, and
  Niu}}]{Xiao:2010a}
\bibinfo{author}{\bibfnamefont{D.}~\bibnamefont{Xiao}},
  \bibinfo{author}{\bibfnamefont{M.-C.} \bibnamefont{Chang}}, \bibnamefont{and}
  \bibinfo{author}{\bibfnamefont{Q.}~\bibnamefont{Niu}}, \bibinfo{journal}{Rev.
  Mod. Phys.} \textbf{\bibinfo{volume}{82}}, \bibinfo{pages}{1959}
  (\bibinfo{year}{2010}),
  \urlprefix\url{http://link.aps.org/doi/10.1103/RevModPhys.82.1959}.

\bibitem[{\citenamefont{Ochoa and Rold\'an}(2013)}]{Ochoa:2013a}
\bibinfo{author}{\bibfnamefont{H.}~\bibnamefont{Ochoa}} \bibnamefont{and}
  \bibinfo{author}{\bibfnamefont{R.}~\bibnamefont{Rold\'an}},
  \bibinfo{journal}{Phys. Rev. B} \textbf{\bibinfo{volume}{87}},
  \bibinfo{pages}{245421} (\bibinfo{year}{2013}),
  \urlprefix\url{http://link.aps.org/doi/10.1103/PhysRevB.87.245421}.

\bibitem[{\citenamefont{Wang and Wu}(2013)}]{Wang:2013b}
\bibinfo{author}{\bibfnamefont{L.}~\bibnamefont{Wang}} \bibnamefont{and}
  \bibinfo{author}{\bibfnamefont{M.~W.} \bibnamefont{Wu}},
  \bibinfo{journal}{e-print} \textbf{\bibinfo{volume}{arXiv:1305.3361}}
  (\bibinfo{year}{2013}).

\bibitem[{\citenamefont{Mak et~al.}(2012)\citenamefont{Mak, He, Shan, and
  Heinz}}]{Mak:2012a}
\bibinfo{author}{\bibfnamefont{K.~F.} \bibnamefont{Mak}},
  \bibinfo{author}{\bibfnamefont{K.}~\bibnamefont{He}},
  \bibinfo{author}{\bibfnamefont{J.}~\bibnamefont{Shan}}, \bibnamefont{and}
  \bibinfo{author}{\bibfnamefont{T.~F.} \bibnamefont{Heinz}},
  \bibinfo{journal}{Nat. Nanotechnol.} \textbf{\bibinfo{volume}{7}},
  \bibinfo{pages}{494} (\bibinfo{year}{2012}).

\bibitem[{\citenamefont{Zeng et~al.}(2012)\citenamefont{Zeng, Dai, Yao, Xiao,
  and Cui}}]{Zeng:2012a}
\bibinfo{author}{\bibfnamefont{H.}~\bibnamefont{Zeng}},
  \bibinfo{author}{\bibfnamefont{J.}~\bibnamefont{Dai}},
  \bibinfo{author}{\bibfnamefont{W.}~\bibnamefont{Yao}},
  \bibinfo{author}{\bibfnamefont{D.}~\bibnamefont{Xiao}}, \bibnamefont{and}
  \bibinfo{author}{\bibfnamefont{X.}~\bibnamefont{Cui}}, \bibinfo{journal}{Nat.
  Nanotechnol.} \textbf{\bibinfo{volume}{7}}, \bibinfo{pages}{490}
  (\bibinfo{year}{2012}).

\bibitem[{\citenamefont{Sallen et~al.}(2012)\citenamefont{Sallen, Bouet, Marie,
  Wang, Zhu, Han, Lu, Tan, Amand, Liu et~al.}}]{Sallen:2012a}
\bibinfo{author}{\bibfnamefont{G.}~\bibnamefont{Sallen}},
  \bibinfo{author}{\bibfnamefont{L.}~\bibnamefont{Bouet}},
  \bibinfo{author}{\bibfnamefont{X.}~\bibnamefont{Marie}},
  \bibinfo{author}{\bibfnamefont{G.}~\bibnamefont{Wang}},
  \bibinfo{author}{\bibfnamefont{C.~R.} \bibnamefont{Zhu}},
  \bibinfo{author}{\bibfnamefont{W.~P.} \bibnamefont{Han}},
  \bibinfo{author}{\bibfnamefont{Y.}~\bibnamefont{Lu}},
  \bibinfo{author}{\bibfnamefont{P.~H.} \bibnamefont{Tan}},
  \bibinfo{author}{\bibfnamefont{T.}~\bibnamefont{Amand}},
  \bibinfo{author}{\bibfnamefont{B.~L.} \bibnamefont{Liu}},
  \bibnamefont{et~al.}, \bibinfo{journal}{Phys. Rev. B}
  \textbf{\bibinfo{volume}{86}}, \bibinfo{pages}{081301}
  (\bibinfo{year}{2012}),
  \urlprefix\url{http://link.aps.org/doi/10.1103/PhysRevB.86.081301}.

\bibitem[{\citenamefont{Kioseoglou et~al.}(2012)\citenamefont{Kioseoglou,
  Hanbicki, Currie, Friedman, Gunlycke, and Jonker}}]{Kioseoglou:2012a}
\bibinfo{author}{\bibfnamefont{G.}~\bibnamefont{Kioseoglou}},
  \bibinfo{author}{\bibfnamefont{A.~T.} \bibnamefont{Hanbicki}},
  \bibinfo{author}{\bibfnamefont{M.}~\bibnamefont{Currie}},
  \bibinfo{author}{\bibfnamefont{A.~L.} \bibnamefont{Friedman}},
  \bibinfo{author}{\bibfnamefont{D.}~\bibnamefont{Gunlycke}}, \bibnamefont{and}
  \bibinfo{author}{\bibfnamefont{B.~T.} \bibnamefont{Jonker}},
  \bibinfo{journal}{Applied Physics Letters} \textbf{\bibinfo{volume}{101}},
  \bibinfo{eid}{221907} (pages~\bibinfo{numpages}{4}) (\bibinfo{year}{2012}),
  \urlprefix\url{http://link.aip.org/link/?APL/101/221907/1}.

\bibitem[{\citenamefont{Wu et~al.}(2013)\citenamefont{Wu, Huang, Aivazian,
  Ross, Cobden, and Xu}}]{Wu:2013a}
\bibinfo{author}{\bibfnamefont{S.}~\bibnamefont{Wu}},
  \bibinfo{author}{\bibfnamefont{C.}~\bibnamefont{Huang}},
  \bibinfo{author}{\bibfnamefont{G.}~\bibnamefont{Aivazian}},
  \bibinfo{author}{\bibfnamefont{J.~S.} \bibnamefont{Ross}},
  \bibinfo{author}{\bibfnamefont{D.~H.} \bibnamefont{Cobden}},
  \bibnamefont{and} \bibinfo{author}{\bibfnamefont{X.}~\bibnamefont{Xu}},
  \bibinfo{journal}{ACS Nano} \textbf{\bibinfo{volume}{7}},
  \bibinfo{pages}{2768} (\bibinfo{year}{2013}),
  \eprint{http://pubs.acs.org/doi/pdf/10.1021/nn4002038},
  \urlprefix\url{http://pubs.acs.org/doi/abs/10.1021/nn4002038}.

\bibitem[{\citenamefont{Meier and Zakharchenya}(1984)}]{Meier:1984a}
\bibinfo{author}{\bibfnamefont{F.}~\bibnamefont{Meier}} \bibnamefont{and}
  \bibinfo{author}{\bibfnamefont{B.}~\bibnamefont{Zakharchenya}},
  \bibinfo{journal}{Modern Problems in Condensed Matter Sciences
  (North-Holland, Amsterdam).} \textbf{\bibinfo{volume}{8}}
  (\bibinfo{year}{1984}).

\bibitem[{\citenamefont{Novoselov et~al.}(2005)\citenamefont{Novoselov, Jiang,
  Schedin, Booth, Khotkevich, Morozov, and Geim}}]{Novoselov:2005a}
\bibinfo{author}{\bibfnamefont{K.~S.} \bibnamefont{Novoselov}},
  \bibinfo{author}{\bibfnamefont{D.}~\bibnamefont{Jiang}},
  \bibinfo{author}{\bibfnamefont{F.}~\bibnamefont{Schedin}},
  \bibinfo{author}{\bibfnamefont{T.~J.} \bibnamefont{Booth}},
  \bibinfo{author}{\bibfnamefont{V.~V.} \bibnamefont{Khotkevich}},
  \bibinfo{author}{\bibfnamefont{S.~V.} \bibnamefont{Morozov}},
  \bibnamefont{and} \bibinfo{author}{\bibfnamefont{A.~K.} \bibnamefont{Geim}},
  \bibinfo{journal}{Proc. Natl Acad. Sci. USA} \textbf{\bibinfo{volume}{102}},
  \bibinfo{pages}{10451} (\bibinfo{year}{2005}).

\bibitem[{\citenamefont{Sallen et~al.}(2011)\citenamefont{Sallen, Urbaszek,
  Glazov, Ivchenko, Kuroda, Mano, Kunz, Abbarchi, Sakoda, Lagarde
  et~al.}}]{Sallen:2011a}
\bibinfo{author}{\bibfnamefont{G.}~\bibnamefont{Sallen}},
  \bibinfo{author}{\bibfnamefont{B.}~\bibnamefont{Urbaszek}},
  \bibinfo{author}{\bibfnamefont{M.~M.} \bibnamefont{Glazov}},
  \bibinfo{author}{\bibfnamefont{E.~L.} \bibnamefont{Ivchenko}},
  \bibinfo{author}{\bibfnamefont{T.}~\bibnamefont{Kuroda}},
  \bibinfo{author}{\bibfnamefont{T.}~\bibnamefont{Mano}},
  \bibinfo{author}{\bibfnamefont{S.}~\bibnamefont{Kunz}},
  \bibinfo{author}{\bibfnamefont{M.}~\bibnamefont{Abbarchi}},
  \bibinfo{author}{\bibfnamefont{K.}~\bibnamefont{Sakoda}},
  \bibinfo{author}{\bibfnamefont{D.}~\bibnamefont{Lagarde}},
  \bibnamefont{et~al.}, \bibinfo{journal}{Phys. Rev. Lett.}
  \textbf{\bibinfo{volume}{107}}, \bibinfo{pages}{166604}
  (\bibinfo{year}{2011}).

\bibitem[{apa()}]{apa0}
\bibinfo{note}{The A exciton emission is most likely dominated by the
  negatively charged A$^-$ emission, with a weaker contribution from the
  neutral exciton emission A$^0$. The emission of both complexes is strongly
  polarized due to the chiral optical selection rules in monolayer MoS$_2$
  \cite{Mak:2013a}.}

\bibitem[{\citenamefont{Korn et~al.}(2011)\citenamefont{Korn, Heydrich, Hirmer,
  Schmutzler, and Sch\"{u}ller}}]{Korn:2011a}
\bibinfo{author}{\bibfnamefont{T.}~\bibnamefont{Korn}},
  \bibinfo{author}{\bibfnamefont{S.}~\bibnamefont{Heydrich}},
  \bibinfo{author}{\bibfnamefont{M.}~\bibnamefont{Hirmer}},
  \bibinfo{author}{\bibfnamefont{J.}~\bibnamefont{Schmutzler}},
  \bibnamefont{and}
  \bibinfo{author}{\bibfnamefont{C.}~\bibnamefont{Sch\"{u}ller}},
  \bibinfo{journal}{Applied Physics Letters} \textbf{\bibinfo{volume}{99}},
  \bibinfo{eid}{102109} (\bibinfo{year}{2011}).

\bibitem[{\citenamefont{Shi et~al.}(2013)\citenamefont{Shi, Yan, Bertolazzi,
  Brivio, Gao, Kis, Jena, Xing, and Huang}}]{Shi:2013b}
\bibinfo{author}{\bibfnamefont{H.}~\bibnamefont{Shi}},
  \bibinfo{author}{\bibfnamefont{R.}~\bibnamefont{Yan}},
  \bibinfo{author}{\bibfnamefont{S.}~\bibnamefont{Bertolazzi}},
  \bibinfo{author}{\bibfnamefont{J.}~\bibnamefont{Brivio}},
  \bibinfo{author}{\bibfnamefont{B.}~\bibnamefont{Gao}},
  \bibinfo{author}{\bibfnamefont{A.}~\bibnamefont{Kis}},
  \bibinfo{author}{\bibfnamefont{D.}~\bibnamefont{Jena}},
  \bibinfo{author}{\bibfnamefont{H.~G.} \bibnamefont{Xing}}, \bibnamefont{and}
  \bibinfo{author}{\bibfnamefont{L.}~\bibnamefont{Huang}},
  \bibinfo{journal}{ACS Nano} \textbf{\bibinfo{volume}{7}},
  \bibinfo{pages}{1072} (\bibinfo{year}{2013}),
  \eprint{http://pubs.acs.org/doi/pdf/10.1021/nn303973r},
  \urlprefix\url{http://pubs.acs.org/doi/abs/10.1021/nn303973r}.

\bibitem[{\citenamefont{Korm\'anyos et~al.}(2013)\citenamefont{Korm\'anyos,
  Z\'olyomi, Drummond, Rakyta, Burkard, and Fal'ko}}]{Kormanyos:2013a}
\bibinfo{author}{\bibfnamefont{A.}~\bibnamefont{Korm\'anyos}},
  \bibinfo{author}{\bibfnamefont{V.}~\bibnamefont{Z\'olyomi}},
  \bibinfo{author}{\bibfnamefont{N.~D.} \bibnamefont{Drummond}},
  \bibinfo{author}{\bibfnamefont{P.}~\bibnamefont{Rakyta}},
  \bibinfo{author}{\bibfnamefont{G.}~\bibnamefont{Burkard}}, \bibnamefont{and}
  \bibinfo{author}{\bibfnamefont{V.~I.} \bibnamefont{Fal'ko}},
  \bibinfo{journal}{Phys. Rev. B} \textbf{\bibinfo{volume}{88}},
  \bibinfo{pages}{045416} (\bibinfo{year}{2013}),
  \urlprefix\url{http://link.aps.org/doi/10.1103/PhysRevB.88.045416}.

\bibitem[{\citenamefont{Zhu et~al.}(2013)\citenamefont{Zhu, Wang, Liu, Marie,
  Qiao, Zhang, Wu, Fan, Tan, Amand et~al.}}]{Wang:2013c}
\bibinfo{author}{\bibfnamefont{C.~R.} \bibnamefont{Zhu}},
  \bibinfo{author}{\bibfnamefont{G.}~\bibnamefont{Wang}},
  \bibinfo{author}{\bibfnamefont{B.~L.} \bibnamefont{Liu}},
  \bibinfo{author}{\bibfnamefont{X.}~\bibnamefont{Marie}},
  \bibinfo{author}{\bibfnamefont{X.~F.} \bibnamefont{Qiao}},
  \bibinfo{author}{\bibfnamefont{X.}~\bibnamefont{Zhang}},
  \bibinfo{author}{\bibfnamefont{X.~X.} \bibnamefont{Wu}},
  \bibinfo{author}{\bibfnamefont{H.}~\bibnamefont{Fan}},
  \bibinfo{author}{\bibfnamefont{P.~H.} \bibnamefont{Tan}},
  \bibinfo{author}{\bibfnamefont{T.}~\bibnamefont{Amand}},
  \bibnamefont{et~al.}, \bibinfo{journal}{Phys. Rev. B}
  \textbf{\bibinfo{volume}{88}}, \bibinfo{pages}{121301}
  (\bibinfo{year}{2013}).

\bibitem[{pol()}]{pol300K}
\bibinfo{note}{Substantial broadening, and hence overlap of the transitions at
  300K can lower the efficiency of valley polarization generation during
  absorption.}

\bibitem[{\citenamefont{Wang et~al.}(2012{\natexlab{b}})\citenamefont{Wang,
  Ruzicka, Kumar, Bellus, Chiu, and Zhao}}]{Wang:2012b}
\bibinfo{author}{\bibfnamefont{R.}~\bibnamefont{Wang}},
  \bibinfo{author}{\bibfnamefont{B.~A.} \bibnamefont{Ruzicka}},
  \bibinfo{author}{\bibfnamefont{N.}~\bibnamefont{Kumar}},
  \bibinfo{author}{\bibfnamefont{M.~Z.} \bibnamefont{Bellus}},
  \bibinfo{author}{\bibfnamefont{H.-Y.} \bibnamefont{Chiu}}, \bibnamefont{and}
  \bibinfo{author}{\bibfnamefont{H.}~\bibnamefont{Zhao}},
  \bibinfo{journal}{Phys. Rev. B} \textbf{\bibinfo{volume}{86}},
  \bibinfo{pages}{045406} (\bibinfo{year}{2012}{\natexlab{b}}),
  \urlprefix\url{http://link.aps.org/doi/10.1103/PhysRevB.86.045406}.

\bibitem[{\citenamefont{Varene et~al.}(2012)\citenamefont{Varene, Bogner,
  Bronner, and Tegeder}}]{Varene:2012a}
\bibinfo{author}{\bibfnamefont{E.}~\bibnamefont{Varene}},
  \bibinfo{author}{\bibfnamefont{L.}~\bibnamefont{Bogner}},
  \bibinfo{author}{\bibfnamefont{C.}~\bibnamefont{Bronner}}, \bibnamefont{and}
  \bibinfo{author}{\bibfnamefont{P.}~\bibnamefont{Tegeder}},
  \bibinfo{journal}{Phys. Rev. Lett.} \textbf{\bibinfo{volume}{109}},
  \bibinfo{pages}{207601} (\bibinfo{year}{2012}),
  \urlprefix\url{http://link.aps.org/doi/10.1103/PhysRevLett.109.207601}.

\bibitem[{\citenamefont{Perebeinos et~al.}(2005)\citenamefont{Perebeinos,
  Tersoff, and Avouris}}]{Perebeinos:2005a}
\bibinfo{author}{\bibfnamefont{V.}~\bibnamefont{Perebeinos}},
  \bibinfo{author}{\bibfnamefont{J.}~\bibnamefont{Tersoff}}, \bibnamefont{and}
  \bibinfo{author}{\bibfnamefont{P.}~\bibnamefont{Avouris}},
  \bibinfo{journal}{Nano Letters} \textbf{\bibinfo{volume}{5}},
  \bibinfo{pages}{2495} (\bibinfo{year}{2005}),
  \eprint{http://pubs.acs.org/doi/pdf/10.1021/nl051828s},
  \urlprefix\url{http://pubs.acs.org/doi/abs/10.1021/nl051828s}.

\bibitem[{\citenamefont{Watanabe et~al.}(1997)\citenamefont{Watanabe, Asahi,
  Fukumura, Masuhara, Hamano, and Kurata}}]{Watanabe:1997a}
\bibinfo{author}{\bibfnamefont{K.}~\bibnamefont{Watanabe}},
  \bibinfo{author}{\bibfnamefont{T.}~\bibnamefont{Asahi}},
  \bibinfo{author}{\bibfnamefont{H.}~\bibnamefont{Fukumura}},
  \bibinfo{author}{\bibfnamefont{H.}~\bibnamefont{Masuhara}},
  \bibinfo{author}{\bibfnamefont{K.}~\bibnamefont{Hamano}}, \bibnamefont{and}
  \bibinfo{author}{\bibfnamefont{T.}~\bibnamefont{Kurata}},
  \bibinfo{journal}{The Journal of Physical Chemistry B}
  \textbf{\bibinfo{volume}{101}}, \bibinfo{pages}{1510} (\bibinfo{year}{1997}),
  \eprint{http://pubs.acs.org/doi/pdf/10.1021/jp9609287},
  \urlprefix\url{http://pubs.acs.org/doi/abs/10.1021/jp9609287}.

\bibitem[{\citenamefont{Feldmann et~al.}(1987)\citenamefont{Feldmann, Peter,
  G\"obel, Dawson, Moore, Foxon, and Elliott}}]{Feldmann:1987a}
\bibinfo{author}{\bibfnamefont{J.}~\bibnamefont{Feldmann}},
  \bibinfo{author}{\bibfnamefont{G.}~\bibnamefont{Peter}},
  \bibinfo{author}{\bibfnamefont{E.~O.} \bibnamefont{G\"obel}},
  \bibinfo{author}{\bibfnamefont{P.}~\bibnamefont{Dawson}},
  \bibinfo{author}{\bibfnamefont{K.}~\bibnamefont{Moore}},
  \bibinfo{author}{\bibfnamefont{C.}~\bibnamefont{Foxon}}, \bibnamefont{and}
  \bibinfo{author}{\bibfnamefont{R.~J.} \bibnamefont{Elliott}},
  \bibinfo{journal}{Phys. Rev. Lett.} \textbf{\bibinfo{volume}{59}},
  \bibinfo{pages}{2337} (\bibinfo{year}{1987}),
  \urlprefix\url{http://link.aps.org/doi/10.1103/PhysRevLett.59.2337}.

\bibitem[{\citenamefont{Jones et~al.}(2013)}]{Jones:2013a}
\bibinfo{author}{\bibfnamefont{A.}~\bibnamefont{Jones}} \bibnamefont{et~al.},
  \bibinfo{journal}{Nat. Nanotechnol.} \textbf{\bibinfo{volume}{8}},
  \bibinfo{pages}{634} (\bibinfo{year}{2013}).

\bibitem[{\citenamefont{Urbaszek et~al.}(2013)\citenamefont{Urbaszek, Marie,
  Amand, Krebs, Voisin, Maletinsky, H\"ogele, and Imamoglu}}]{Urbaszek:2013a}
\bibinfo{author}{\bibfnamefont{B.}~\bibnamefont{Urbaszek}},
  \bibinfo{author}{\bibfnamefont{X.}~\bibnamefont{Marie}},
  \bibinfo{author}{\bibfnamefont{T.}~\bibnamefont{Amand}},
  \bibinfo{author}{\bibfnamefont{O.}~\bibnamefont{Krebs}},
  \bibinfo{author}{\bibfnamefont{P.}~\bibnamefont{Voisin}},
  \bibinfo{author}{\bibfnamefont{P.}~\bibnamefont{Maletinsky}},
  \bibinfo{author}{\bibfnamefont{A.}~\bibnamefont{H\"ogele}}, \bibnamefont{and}
  \bibinfo{author}{\bibfnamefont{A.}~\bibnamefont{Imamoglu}},
  \bibinfo{journal}{Rev. Mod. Phys.} \textbf{\bibinfo{volume}{85}},
  \bibinfo{pages}{79} (\bibinfo{year}{2013}),
  \urlprefix\url{http://link.aps.org/doi/10.1103/RevModPhys.85.79}.

\end{thebibliography}
\end{document}